# Electronic properties of Janus black arsenic-phosphorus (b-AsP) nanoribbons under transverse electric field

Zhongqi Ren[a1], Chen Li[a1], Xinnuo Huang[a], Xuefei Yan[a,b] , Weiqi Li[a]*

*a School of Physics, Harbin Institute of Technology, Harbin 150001, China*
*b School of Microelectronics Science and Technology, Sun Yat-Sen University,*
*Zhuhai 519082, People's Republic of China*

## Abstract

The electronic transport properties of Janus monolayer black arsenic phosphorus (b-AsP) nanoribbons have been investigated utilizing the tight-binding approach. The dependence of electronic structure on edge structures is systematically investigated. (1,3)nb and (3,1)nb b-AsP nanoribbons exhibit flatter edge bands than zigzag and armchair counterparts. The edge band of the armchair ones show double degeneracy. Further, the calculated results show the band gap of the zigzag ribbon with different boundary morphology is widely tunable by transverse electric field. A critical electric field can fully close the gap and induce a phase transition from a semiconductor into a conductor. Our work suggests dynamically tunable bandgap in Janus b-AsP nanoribbons and reveals the potential of Janus b-AsP for transmission devices.

## 1.Introduction

Since graphene offers new inroads into materials science and condensed-matter physics [1], numerous graphene-related materials (GRM) have been researched in recent decades. Among the 2D materials, monolayer black phosphorus (BP) has aroused wide concern due to its dramatic anisotropy [2–8]. As the most stable allotrope [9,10], BP is a van der Waals bonded layered material similar to graphite, so that its monolayer can be fabricated by mechanical exfoliation [2,8,11–16]. Researches show that the band gap of BP can be tuned from 0.3 eV (bulk form) to 1.7 eV (monolayer) [17–19]. Hence, the monolayer BP with sizable band gap obtains a strong capability to absorb mid-wavelength infrared regime (MWIR) [20]. electrical conductivity and optical responses due to its anisotropic band structure, [2,6,8,11,21] gives huge promise for applications in field effect transistors (FET) and optoelectronic devices (OPD) [22].

Despite the great potential for applications in electronics and optoelectronics, BP's degradation in the presence of atomic vacancies [23] and ambient oxygen molecules [24] is an existing obstacle to resolve[25–27] and the search for BP-derived materials has been hot topics. Black arsenic–phosphorus (b-AsP), as an alloy of BP with arsenic atoms, retains the semiconducting nature with anisotropy in electronic and optical properties [28]. The tunable bandgaps in the long-wavelength infrared regime (LWIR), which can be modulated from 0.15 to 0.3 eV, are scarce for layered materials [28,29]. Moreover, detectors with good responsiveness based on b-AsP have been manufactured [20,29,30]. Recently, Li et al predicted a new type of monolayer b-AsP with Janus structure which is totally different from the alloy form [31]. The Janus monolayer b-AsP has a puckered structure with similarity to BP. It consists of two sublayers, i.e. As- and P-sublayer, which generates additional out-of-plane asymmetry. This interesting 2D material has been proved stable with strong optical and elastic anisotropic and expected to be synthesized by Janus engineering and alloying methods the same way as other Janus 2D crystals

[31–34]. Moreover, quantum dots consisting of Janus monolayer b-AsP has been discovered a tunable optical absorption under a perpendicular magnetic field [35]. More electronic and optical properties of this novel 2D material are waiting for exploration.

Recently, some nanoribbons with various edge atomic configurations have been studied widely. Due to quantum confinement effects[36], the edge geometry can result in an abundant electrical property. For example, the effects of chirality and edge geometry on the anisotropic thermal transport properties have been studied for $MoS_2$ nanoribbons. Ribbons with chiralities other than 0° and 30° always have lower thermal conductivity, yet a local maximum at 19.1° [37]. Black phosphorus and alloy b-AsP nanoribbons with beard-zigzag edge [7], skewed-zigzag edge (SZ), and skewed-armchair edge(SA) have also been reported [38 – 40] which display different electronic properties [41]. However, the exploration of b-AsP nanoribbons with different edges have not been studied so far.

In this work, we focus on the electronic properties of monolayer Janus b-AsP nanoribbons with various edge geometries and varies under a transverse electric field. The rest of this article is organized as follows. Section 2 introduces the tight-binding models of b-AsP and the formula used for calculation. In section 3, we separately analyze the influence of electric field on electronic structure of zigzag/armchair b-AsP nanoribbons, as well as b-AsP nanoribbons with chiral angles. Finally, we summarize the main conclusions in section 4.

## 2. Model and formulation

Janus b-AsP nanoribbons have a low-puckered honeycomb surface. Each P/As atom away from the edges is covalently bonded with two adjacent P/As atoms and one As/P atom, forming a two-sublayer structure as shown in figure 1. Each sublayer is completely composed of the same kind of atoms different from another sublayer, which generates additional out-of-plane asymmetry. The three different bond lengths are listed in Table I.

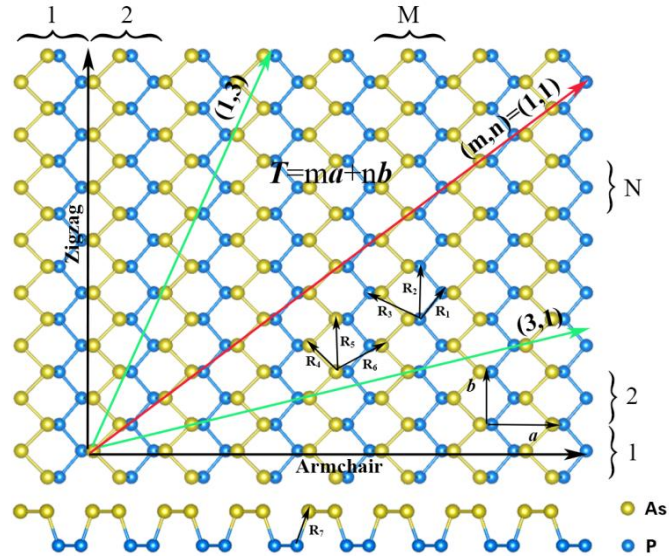

Figure 1: Top view and side view of monolayer b-AsP. $T$ =m$\boldsymbol{a}$ + n$\boldsymbol{b}$ denote the chiral vector. The vectors $R_1$, $R_2$ and $R_3$ denote the intralayer bonds between P atoms while $R_4$, $R_5$ and $R_6$ denote those between As atoms.

The vector $R_7$ denotes the interlayer bonds between P and As atoms.

| $a$ (nm) | $b$ (nm) | $d_{P-P}(R_1)$ (nm) | $d_{As-As}(R_4)$ (nm) | $d_{P-As}(R_7)$ (nm) |
|---|---|---|---|---|
| 4.71 | 3.67 | 2.28 | 2.45 | 2.38 |

Table 1: lattice parameters of Janus b-AsP structure including lattice constants and bond lengths.

We define the number of unit cells along the armchair/zigzag direction as the width N/M of the zigzag/armchair ribbon illustrated in figure 1. By tailoring a Janus b-AsP sheet along the two principal directions, i.e., the armchair and zigzag directions, we can obtain the atomic models of the armchair/zigzag Janus b-AsP nanoribbons as shown in figure 2(a/b). With the optimized lattice constants with a = 4.72 ˚A and b = 3.47 ˚A [31], each rectangular unit cell contains two P atoms and two As atoms.

To describe the skewed edges, we can define a chiral vector $T$, which is given by

$$T = m\boldsymbol{a} + n\boldsymbol{b}$$

where $\boldsymbol{a}$ and $\boldsymbol{b}$ are the primitive vectors of b-AsP. Moreover, according to the number of dangled bonds of the outermost atoms, the skewed edges can be divided into two categories: bearded and non-bearded. For bearded (non-bearded) boundary edges, the number of dangled bonds of the outermost atoms is two (one) per atom [42], as shown in Figure 2. For black phosphorus, the stability for (3,1) PNR and (1,1) PNR has been studied in previous work[ref] and it has been shown that the non-bearded ribbon is more stable. The stability of b-AsP ribbons has not been studied yet, but generally, the non-bearded(nb) ribbon is usually more stable than bearded(b) ribbons, for they have fewer dangled bonds of the outermost atoms. In this paper, we calculated the electronic properties (1,3) nb AsP ribbon, (3,1) nb AsP ribbon as well as AsP ribbon with zigzag/armchair edges.

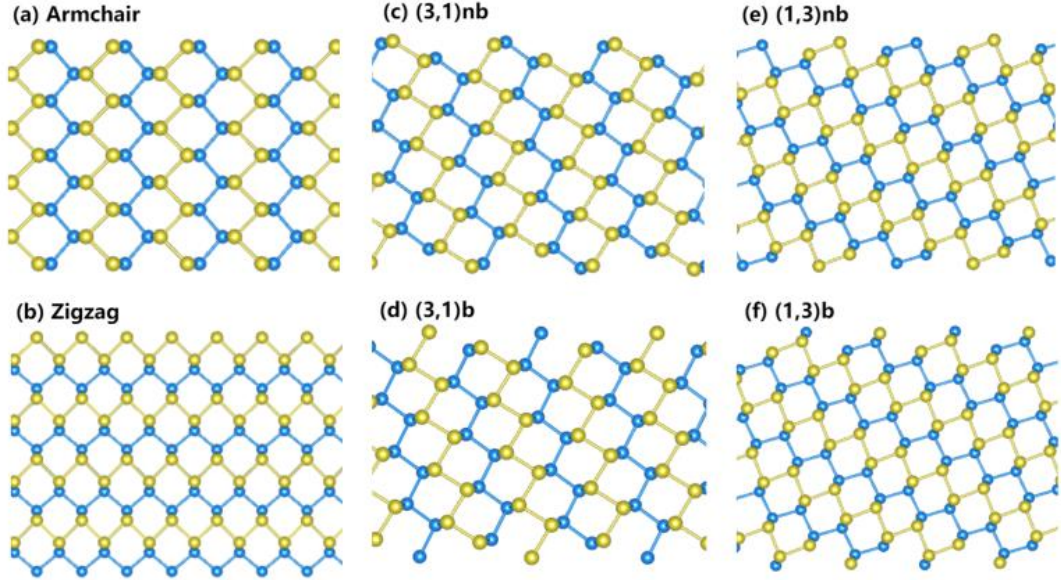

Figure 2: illustrations of (a) (3,1)nbPNR, (b) (1,1)bPNR, (c) (1,3)bPNR, (d) (3,1)bPNR, (e) (1,1)nbPNR, and (f) (1,3)nbPNR, where the red-dashed-line parallelogram in each PNR indicates its minimum periodical supercell.

The low-energy electronic structure of monolayer Janus b-AsP can be described by a multiband tight-binding (TB) model. The Hamiltonian for the Janus b-AsP nanoribbons is defined as:

$$H = \sum_{i,\mu\nu} \varepsilon_{\mu,\nu} c_{i,\mu}^{\dagger} c_{i,\nu} + \sum_{(i,j),\mu\nu} [t_{ij,\mu\nu} c_{i,\mu}^{\dagger} c_{j,\nu} + h.c.], \tag{1}$$

where $c_{i,\mu}^{\dagger}$ ($c_{i,\nu}$) is the creation (annihilation) operator of the electron at the site $i$ in the atomic orbital $\mu$, including the basis of sp³ orbitals, i.e., $s$, $p_x$, $p_y$ and $p_z$. The summation $(i,j)$ is taken if the bond of the two sites $i, j$ is accord with any of the seven types of bonds denoted by the vectors in figure 1.

Utilizing the Slater-Koster (SK) tight-binding (TB) method [31,36], we calculated the off-diagonal matrices of all the different bonds containing hopping integrals in the basis of sp³ orbitals as follows. Specific proof is shown in support information.

$$h_1(R_1) = \begin{pmatrix} -2.11 & 3.09 & 0 & -2.62 \\ -3.09 & 1.50 & 0 & -2.49 \\ 0 & 0 & -1.44 & 0 \\ 2.62 & -2.49 & 0 & 0.66 \end{pmatrix}, \tag{2}$$

$$h_2(R_1) = \begin{pmatrix} -2.11 & -3.09 & 0 & -2.62 \\ 3.09 & 1.50 & 0 & 2.49 \\ 0 & 0 & -1.44 & 0 \\ 2.62 & 2.49 & 0 & 0.66 \end{pmatrix}, \tag{3}$$

$$h(R_2) = \begin{pmatrix} 0.17 & 0.25 & 0 & 0 \\ -0.25 & -0.50 & 0 & 0 \\ 0 & 0 & -0.01 & 0 \\ 0 & 0 & 0 & -0.01 \end{pmatrix}, \tag{4}$$

$$h_1(R_3) = \begin{pmatrix} -0.50 & 0.16 & 0 & -0.31 \\ -0.16 & -0.07 & 0 & -0.12 \\ 0 & 0 & -0.14 & 0 \\ 0.31 & -0.12 & 0 & 0.09 \end{pmatrix}, \tag{5}$$

$$h_2(R_3) = \begin{pmatrix} -0.50 & -0.16 & 0 & -0.31 \\ 0.16 & -0.07 & 0 & 0.12 \\ 0 & 0 & -0.14 & 0 \\ 0.31 & 0.12 & 0 & 0.09 \end{pmatrix}, \tag{6}$$

$$h_1(R_4) = \begin{pmatrix} -1.22 & 1.84 & 0 & 1.82 \\ -1.84 & 1.28 & 0 & 1.84 \\ 0 & 0 & -0.57 & 0 \\ -1.82 & 1.84 & 0 & 1.26 \end{pmatrix}, \tag{7}$$

$$h_2(R_4) = \begin{pmatrix} -1.22 & -1.84 & 0 & 1.82 \\ 1.84 & 1.28 & 0 & -1.84 \\ 0 & 0 & -0.57 & 0 \\ -1.82 & -1.84 & 0 & 1.26 \end{pmatrix}, \tag{8}$$

$$h(R_5) = \begin{pmatrix} -0.25 & -0.08 & 0 & 0 \\ 0.08 & 0.74 & 0 & 0 \\ 0 & 0 & -0.13 & 0 \\ 0 & 0 & 0 & -0.13 \end{pmatrix}, \tag{9}$$

$$h_1(R_6) = \begin{pmatrix} -0.05 & -0.17 & 0 & -0.29 \\ 0.17 & 0.14 & 0 & 0.66 \\ 0 & 0 & -0.25 & 0 \\ 0.29 & 0.66 & 0 & 0.90 \end{pmatrix}, \tag{10}$$

$$h_2(R_6) = \begin{pmatrix} -0.05 & 0.17 & 0 & -0.29 \\ -0.17 & 0.14 & 0 & -0.66 \\ 0 & 0 & -0.25 & 0 \\ 0.29 & -0.66 & 0 & 0.90 \end{pmatrix}, \tag{11}$$

$$h_1(R_7) = \begin{pmatrix} 1.84 & 0 & 1.16 & 0.39 \\ 0 & 1.02 & 0 & 0 \\ -1.16 & 0 & -2.71 & -1.26 \\ -0.39 & 0 & -1.26 & 0.59 \end{pmatrix}, \tag{12}$$

$$h_2(R_7) = \begin{pmatrix} 1.84 & 0 & -1.16 & 0.39 \\ 0 & 1.02 & 0 & 0 \\ 1.16 & 0 & -2.71 & 1.26 \\ -0.39 & 0 & 1.26 & 0.59 \end{pmatrix}, \tag{13}$$

As seen in Eq (2)~(13), there are two different matrices for all the bonds except $R_2$ and $R_5$. The distinction derives from the chirality of those bonds inducing the opposite numbers at the integrals related to $p_x$ and $p_y$ orbitals.

We can also obtain the diagonal matrices of the onsite energy for P and As atoms.

$$\varepsilon_P = \begin{pmatrix} -8.047 & 0 & 0 & 0 \\ 0 & 1.537 & 0 & 0 \\ 0 & 0 & 1.116 & 0 \\ 0 & 0 & 0 & -2.724 \end{pmatrix}, \tag{14}$$

$$\varepsilon_{As} = \begin{pmatrix} -10.446 & 0 & 0 & 0 \\ 0 & -0.756 & 0 & 0 \\ 0 & 0 & -0.346 & 0 \\ 0 & 0 & 0 & -0.760 \end{pmatrix}, \tag{15}$$

Starting from the Hamiltonian, the local density of states (LDOS) of the system is calculated by the following formula:

$$LDOS(E) = \frac{1}{\Gamma\sqrt{2\pi}} \sum_n |\Psi_n(r)|^2 e^{-(E_n-E)^2/2\Gamma^2}, \tag{16}$$

where $\Gamma$ is a broadening factor and $\Psi_n(r)$ and $E_n$ denoted the eigenfunction and eigenvalue for the $n$ th energy band.

Moreover, the conductance of the nanorribons at zero temperature is determined by the transmission probability and can be calculated using Landauer Büttiker formula [37]:

$$G = \frac{2e^2}{h}T \tag{17}$$

As for the condition at non-zero temperatures, we can utilize the linear response formula:

$$G(E_F) = \frac{2e^2}{h} \int dE[-\partial f(E, E_F)/\partial E]T(E), \tag{18}$$

where $f(E, E_F) = 1/[e^{(E-E_F)/k_BT} + 1]$ is the Fermi-Dirac distribution function and $[-\partial f(E, E_F)/\partial E]$ denotes the thermal broadening function, and $E_F$ is the Fermi energy. Eq (18) satisfies that:

$$T(E) = Tr[\Gamma_L(E,V)G(E,V)\Gamma_R(E,V)G^\dagger(E,V)] \tag{19}$$

where the advanced Green function $G(E,V)/G^\dagger(E,V)$ is the advanced/retarded Green function, and $\Gamma_L(E,V)$ and $\Gamma_R(E,V)$ describe the coupling between the conductor and the left or right leads. Transport properties have been performed by Kwant [43]

## 3. Results and discussion

The calculated band structures and Hall conduction of the zigzag and armchair Janus b-AsP nanoribbons are depicted in figure 3.

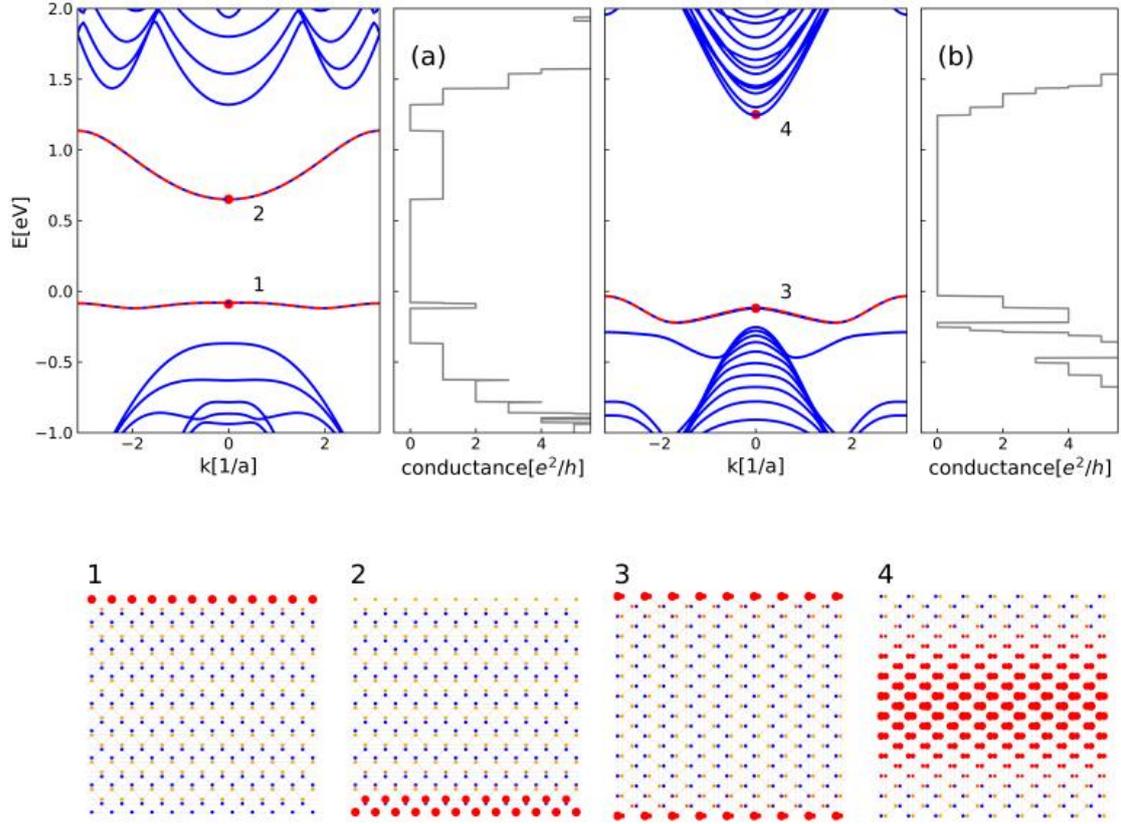

Figure 3: The calculated energy band and conductance for (a) zigzag ribbon, (b) armchair ribbon with widths around 4 nm. The lower panel shows the LDOS at the energy indicated by the red dot in the upper panel.

As seen in figure 3, it is significant to notice the presence of isolated energy bands between the conduction and valence bands for both the armchair and the zigzag nanoribbon. According to the LDOS at four different energies near zero [point 1-4 specified in figure 3], we conclude that these isolated bands are related to the edge states. For the zigzag case, there are two separate bands detaching from the bulk bands, each corresponding to different edge states. As shown in the left panel of figure 3, the conductance exhibits a stepwise structure. The LDOS for the lower band is localized at the upper edge, i.e., the edge terminated by As atoms, while for the upper band the LDOS locates at the edge terminated by P atoms. To simplify the demonstration, we name the upper and the lower edge of the zigzag ribbon after P- and As-terminated edge respectively.

By contrast, we can only detect one isolated energy band close to valence bands of the armchair ribbon. However, the conductance at the corresponding energy is up to 4 units ( $e^2/h$ ), based on the right top panel in figure 3(b), which indicates that this energy band is doubly degenerate. This can also be deduced from the LDOS of point 3 in figure 3, depicting a natural result that the edge states with identical energy in armchair ribbons are localized at both edges, since the armchair ribbon has exactly two same edges. In the following analysis, we will give more valid proof of the degeneracy.

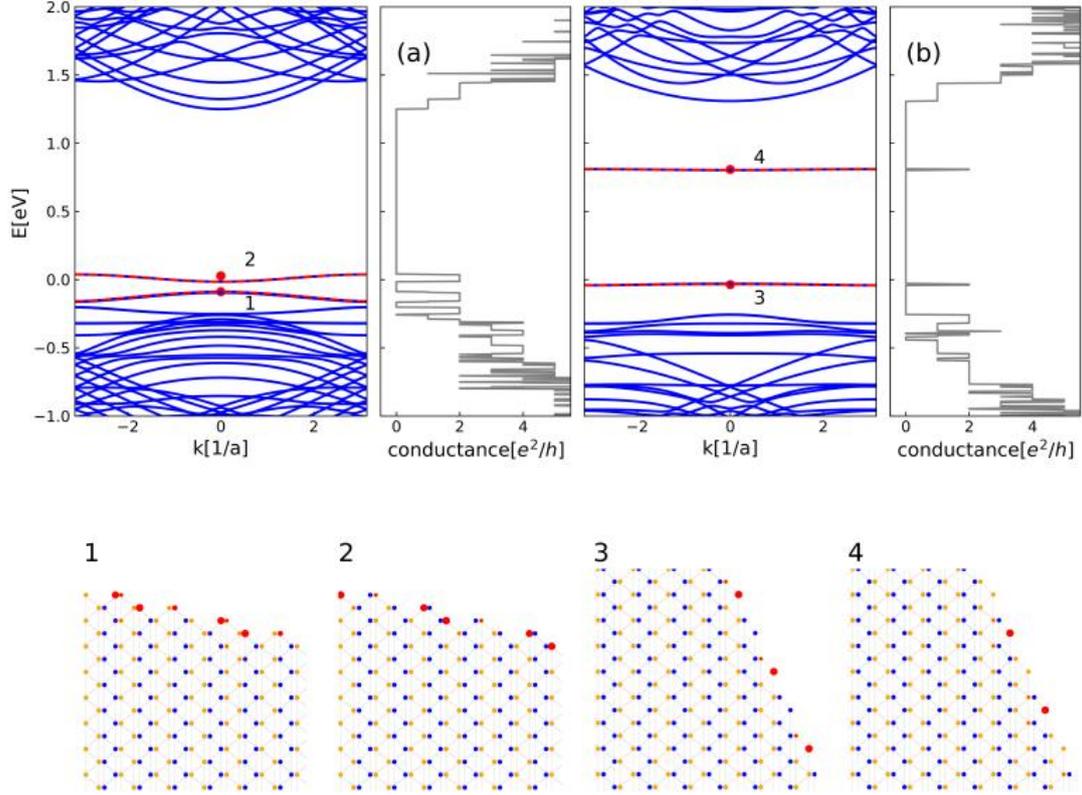

Figure 4: The calculated energy band and conductance for (a) (3,1)nb ribbon, (b) (1,3)nb ribbon with widths around 4 nm. The lower panel shows the LDOS at the energy indicated by the red dot in the upper panel.

For the non-bearded ribbons with the chiral vector (3,1) and (1,3), In consequence of the lone pair electrons at the boundaries, the edge bands are flatter and we can see more localized electron clouds around unsaturated atoms in LDOS. As shown in figure 4(a), there exist two edge bands close to the valence bands in the band structure of the ribbon with the chiral vector (3,1), and the corresponding Hall conductance reduced to 2 units, with LDOS dominated by As and P atoms respectively.

For (1,3)nb ribbons, it is of interest to note that the two edge bands are both quasi-flat, with similar energy of the zigzag edge bands, about -0.04eV for the lower band and 0.8eV for the upper, dominated by As and P atoms respectively. The quasi-flat edge bands also emerge in black phosphorus ribbons [44–46] and their topological origin has been discussed [7].

Energy band of the zigzag ribbon under the transverse electric field were shown in figure 5.

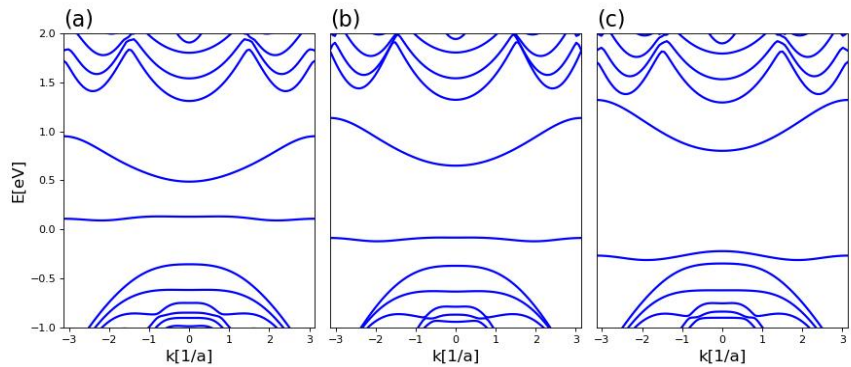

Figure 5: The band structure for the zigzag ribbon under a transverse electric field. (a) refers to the case applied to a positive electric field, (b) the case with zero electric field, and (c) the case with negative field. define the positive direction from As-terminated side to P-terminated side

Apparently, the edge states still exist while the band gap subtly changes. Figure 5(b) refers to the situation without extra field, and figure 5(a) and 5(c) correspond to the band structure under a transverse electronic field $E = +0.1eV / nm$ and $E = -0.1eV / nm$ , respectively. Here, we define the positive direction from P-terminated side to As-terminated side. Combined with the structure of the zigzag nanoribbon in figure 1(b), we derive that the transverse electronic field from the As-terminated edge to the P-terminated edge can close the band gap while the field with opposite direction has a reverse effect. It comes out that the band gap would keep closing until zero if we continue slightly enlarging the magnitude of the field negatively. Figure 6 shows the response of the system under the condition mentioned. In the case for the zigzag nanoribbon around 4-nm wide, the band gap is just closed under the threshold electric field $E = +0.180eV / nm$ , as shown in figure 6(a).

The combined barrier plotted in figure 6(b) by the blue line can be assumed to be derived from the two grey barriers around when the strength of the field reaches the threshold and the gap get exactly closed inducing a phase transition from a semiconductor into a conductor. By further examining the LDOS, the edge states are still present regardless of extra transverse electric field.

Here, we can confirm that the lower edge band in figure 6(a) corresponds to the As-terminated edge state, while the upper one represents the P-terminated edge state. The threshold field modifies the on-site energy of the atoms at edges, causing an energy displacement around $E \cdot width / 2 \sim 0.36eV$ for each edge states. The displacement is almost exactly equivalent to half of the inherent band gap for the zigzag ribbon. And we can see from the LDOS that when the gap closed, the electron transports along two sides at the proper incident energy where the two bands merge. The As-terminated edge get a lot more sector in electronic transport, which could be related to the relatively high conductance of the As-terminated edge state twice as much as that of P-terminated edge.

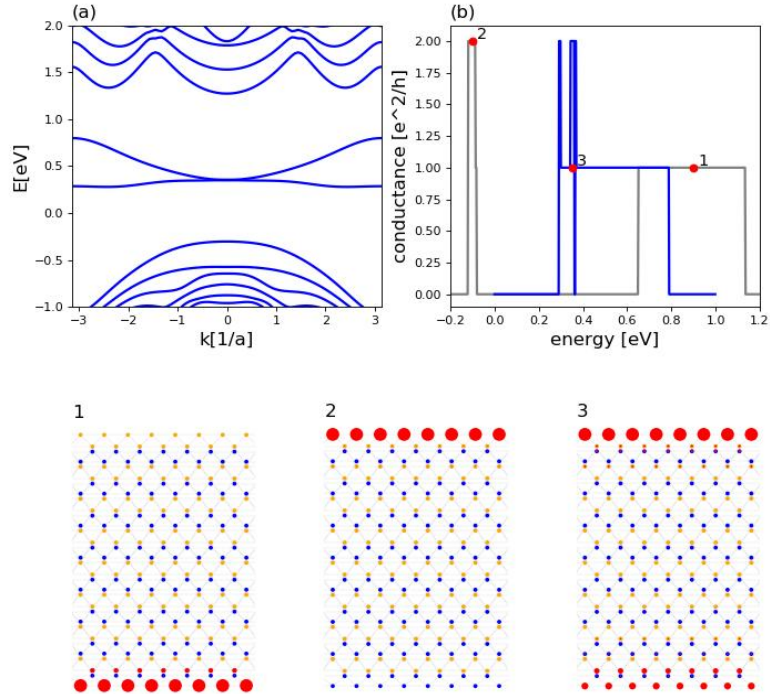

Figure 6: The calculated energy band and conductance for zigzag ribbon under the threshold field. The band gap get closed as shown in (a). the blue line in (b) denotes the change of conductance with the Fermi energy for the system as (a) described, while the grey line denotes the situation without extra electric field. The bottom penal shows the LDOS at the energy indicated by the red dot in (b).(蓝线是多大电场下的电导？)

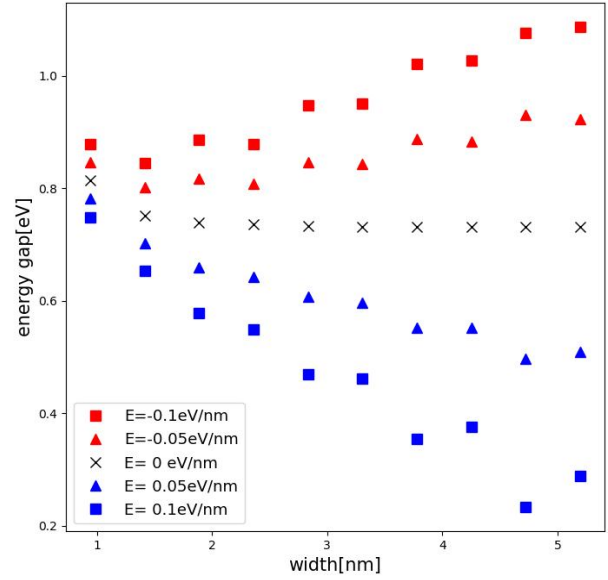

Figure 7: The band gap as the ribbon width varies under extra transverse electric field. Here the election charge 'e' is a dimensionless number.

Moreover, we examined the relation between the gap energy and the width of zigzag nanoribbons under an electric field with different magnitude. As displayed in figure 7, without extra field, the band gap approaches to 0.73 eV as the width grows, which demonstrates that the band gap has little relation to the width and the edge state has a faint size effect. We can see that the wide ribbons are more sensitive to the electric field and the gap and the width are almost linearly correlated within limits. It well illustrates that the effect of the electric field is to change the on-site energy of the atoms at edges so that it can influence the edge states.

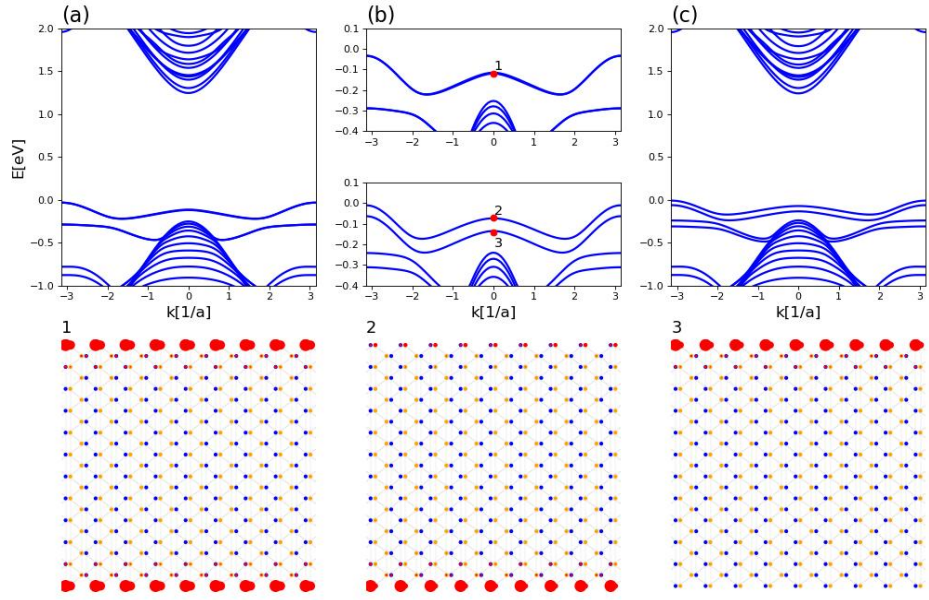

Figure 8: The band structure and LDOS for the armchair ribbon naturally (a) and applied transverse electric field (c). (b) shows the partial enlarged drawing of (a) and (c) around the edge band energy.

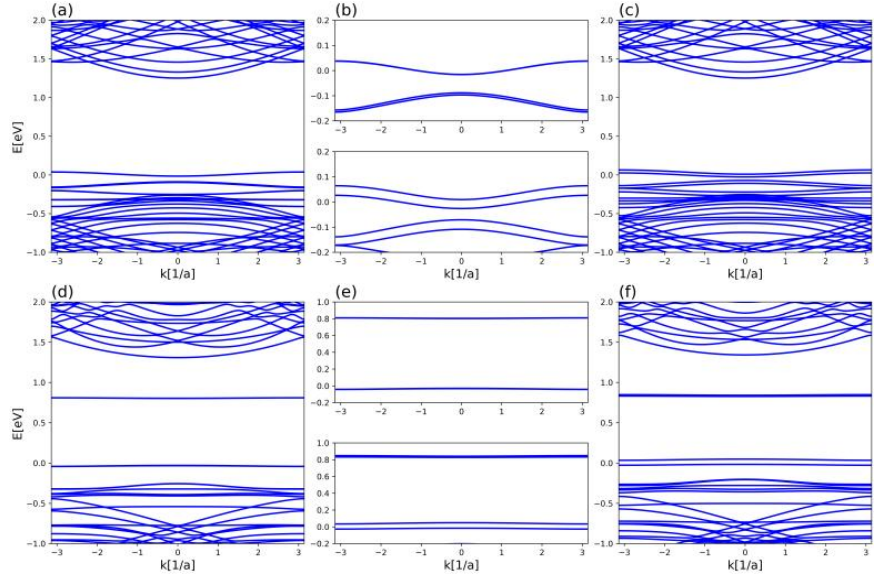

Figure 9: The calculated energy band for the (3,1)nb ribbon (a)/(b)/(c) and the (1,3)nb ribbon (d)/(e)/(f) under transverse electric field. (a)/(d) refer to the natural condition, and (c)/(d) the condition applied the field. (b) and (e) are partial enlarged drawings of (a)/(c) and (d)/(f) respectively.

For armchair nanoribbons and non-bearded ribbons, as shown in figure 8, there are two degenerate energy bands corresponding to the edge states. We can see the LDOS localized at both edges, which differs from the situation in the zigzag ribbon. The reason of the degeneration is that the two edges of the armchair nanoribbons are equivalent and slightly coupled to each other. Through switching the transverse electric field, we can clearly see the twofold-degenerate edge bands split into two separate bands. Ribbons with (3,1)nb and (1,3)nb edges under the transverse electric field also exhibit non-degenerate edge bands, as is shown in figure 9. Apparently, the field changes the asymmetry of the armchair, (3,1)nb and (1,3)nb ribbons and leads to different potential at edges. In addition, we have to mentioned that the zigzag ribbon may have only As-terminated edges or P-terminated edges. For those ribbons, we will also get degenerate bands and the results will be similar to armchair ribbons.

## 4. Summary

We have theoretically studied the electronic properties of the zigzag and armchair b-AsP nanoribbons in the presence of a transverse electric field. By utilizing the $sp^3$ basis set in the tight-binding model, we calculated the band structures and LDOS of the system. The results demonstrate the zigzag ribbon with different edges has a band gap around 0.7 eV between two separate bands close to the zero energy, which corresponds to two edge states each for As and P atoms. This gap is widely tunable by changing the exterior field. A proper field can fully close the gap and induce a phase transition from a semiconductor into a conductor. Depending slightly on the system size, the gap width remains stable. Meanwhile, the armchair ribbon present effectively degenerate edge bands. Applying a transverse electronic field causes the separate non-degenerate edge states and form an energy range of one-edge transport. Similar phenomenon performs as for the condition of (3,1)nb and (1,3)nb ribbons. However, the quasi-flat edge bands emerge in (1,3) nb ribbons and it is a nontrivial result which can be correlated to strongly correlated quantum states.

## (Support information)

Utilizing the Slater-Koster (SK) tight-binding (TB) method [31,36], the hopping integrals $t_{ii,uv}$ can be calculated by the following equations:

$$\langle s|H|p_k\rangle = n_k v_{sp\sigma},$$

where $(v_{ss\sigma}, v_{sp\sigma}, v_{pp\pi}, v_{pp\sigma}, s_{ss\sigma}, s_{sp\sigma}, s_{pp\pi}, s_{pp\sigma})$ is the basis of the directed orbitals in terms of eight integrals fitted from the first-principles calculation [50, 58], $k$ runs over $x$, $y$, and $z$. The directional cosines $n_k = \vec{R} \cdot \vec{e}_k / |\vec{R}|$ is described by the vectors $\vec{R}$ along the bonds as shown in figure 1. Considering the rule of the angular quantum number, the condition $\langle l|H|l'\rangle = (-1)^{l+l'} \langle l'|H|l\rangle$ needs to be satisfied. The Slater-Koster parameters and on-site energies for the structure are shown in Table II. Here, we neglect the slight differences for the atoms on the edge from the others for simplification.

| The Slater–Koster parameters for the bonds in b-AsP [eV] | | | |
|---|---|---|---|
| $\mathbf{R_m}$ | $v_{ss\sigma}$ | $v_{sp\sigma}$ | $v_{pp\sigma}$ | $v_{pp\pi}$ |
| $\mathbf{R_1}$ | −2.114 | −4.053 | 3.607 | −1.442 |
| $\mathbf{R_2}$ | 0.169 | 0.251 | −0.500 | −0.010 |
| $\mathbf{R_3}$ | −0.503 | −0.352 | 0.160 | −0.138 |
| $\mathbf{R_4}$ | −1.219 | −2.587 | 3.107 | −0.570 |
| $\mathbf{R_5}$ | −0.249 | −0.084 | 0.739 | −0.131 |
| $\mathbf{R_6}$ | −0.046 | −0.337 | 1.280 | −0.249 |
| $\mathbf{R_7}$ | 1.839 | 1.226 | −3.133 | 1.017 |

| The on-site parameters for P and As atoms [eV] | | | |
|---|---|---|---|
| Atom | $s$ | $p_x$ | $p_y$ | $p_z$ |
| P | −8.047 | 1.537 | 1.116 | −2.724 |
| As | −10.446 | −0.756 | −0.346 | −0.760 |

# Reference

[1]     Geim A K and Novoselov K S THE RISE OF GRAPHENE 14

[2]     Xia F, Wang H and Jia Y 2014 Rediscovering black phosphorus as an anisotropic layered material for optoelectronics and electronics *Nat Commun* **5** 4458

[3]     Rodin A S, Carvalho A and Castro Neto A H 2014 Strain-Induced Gap Modification in Black Phosphorus *Phys. Rev. Lett.* **112** 176801

[4]     Tran V, Soklaski R, Liang Y and Yang L 2014 Layer-controlled band gap and anisotropic excitons in few-layer black phosphorus *Phys. Rev. B* **89** 235319

[5]     Fei R and Yang L 2014 Strain-Engineering the Anisotropic Electrical Conductance of Few-Layer Black Phosphorus *Nano Lett.* **14** 2884–9

[6]     Qiao J, Kong X, Hu Z-X, Yang F and Ji W 2014 High-mobility transport anisotropy and linear dichroism in few-layer black phosphorus *Nat Commun* **5** 4475

[7]     Ezawa M 2014 Topological origin of quasi-flat edge band in phosphorene *New J. Phys.* **16** 115004

[8]     Liu H, Neal A T, Zhu Z, Luo Z, Xu X, Tománek D and Ye P D 2014 Phosphorene: An Unexplored 2D Semiconductor with a High Hole Mobility *ACS Nano* **8** 4033–41

[9]     Keyes R W 1953 The Electrical Properties of Black Phosphorus *Phys. Rev.* **92** 580–4

[10]    Jamieson J C 1963 Crystal Structures Adopted by Black Phosphorus at High Pressures *Science* **139** 1291–2

[11]    Li L, Yu Y, Ye G J, Ge Q, Ou X, Wu H, Feng D, Chen X H and Zhang Y 2014 Black phosphorus field-effect transistors *Nature Nanotech* **9** 372–7

[12]    Koenig S P, Doganov R A, Schmidt H, Castro Neto A H and Özyilmaz B 2014 Electric field effect in ultrathin black phosphorus *Appl. Phys. Lett.* **104** 103106

[13]    Buscema M, Groenendijk D J, Blanter S I, Steele G A, van der Zant H S J and Castellanos-Gomez A 2014 Fast and Broadband Photoresponse of Few-Layer Black Phosphorus Field-Effect Transistors *Nano Lett.* **14** 3347–52

[14]    Ling X, Wang H, Huang S, Xia F and Dresselhaus M S 2015 The renaissance of black phosphorus *Proc. Natl. Acad. Sci. U.S.A.* **112** 4523–30

[15]    Morita A Semiconducting black phosphorus 16

[16]    Cai Y, Ke Q, Zhang G, Feng Y P, Shenoy V B and Zhang Y-W 2015 Giant Phononic Anisotropy and Unusual Anharmonicity of Phosphorene: Interlayer Coupling and Strain Engineering *Adv. Funct. Mater.* **25** 2230–6

[17]    Han C Q, Yao M Y, Bai X X, Miao L, Zhu F, Guan D D, Wang S, Gao C L, Liu C, Qian D, Liu Y and Jia J 2014 Electronic structure of black phosphorus studied by angle-resolved photoemission spectroscopy *Phys. Rev. B* **90** 085101

[18]    Li L, Kim J, Jin C, Ye G J, Qiu D Y, da Jornada F H, Shi Z, Chen L, Zhang Z, Yang F, Watanabe K, Taniguchi T, Ren W, Louie S G, Chen X H, Zhang Y and Wang F 2017 Direct observation of the layer-dependent electronic structure in phosphorene *Nature Nanotech* **12** 21–5

[19]    Wang X, Jones A M, Seyler K L, Tran V, Jia Y, Zhao H, Wang H, Yang L, Xu X and Xia F 2015 Highly anisotropic and robust excitons in monolayer black phosphorus *Nature Nanotech* **10** 517–21

[20]    Amani M, Regan E, Bullock J, Ahn G H and Javey A 2017 Mid-Wave Infrared Photoconductors Based on Black Phosphorus-Arsenic Alloys *ACS Nano* **11** 11724–31

[21]    Cai Y, Zhang G and Zhang Y-W 2015 Electronic Properties of Phosphorene/Graphene and Phosphorene/Hexagonal Boron Nitride Heterostructures *J. Phys. Chem. C* **119** 13929–36

[22]    Heshmati-Moulai A, Simchi H, Esmaeilzadeh M and Peeters F M 2016 Phase transition and spin-resolved transport in MoS 2 nanoribbons *Phys. Rev. B* **94** 235424

[23]    Cai Y, Ke Q, Zhang G, Yakobson B I and Zhang Y-W 2016 Highly Itinerant Atomic Vacancies in Phosphorene *J. Am. Chem. Soc.* **138** 10199–206

[24]    Kistanov A A, Cai Y, Zhou K, Dmitriev S V and Zhang Y-W 2016 The role of H $_2$ O and O $_2$ molecules and phosphorus vacancies in the structure instability of phosphorene *2D Mater.* **4** 015010

[25]    Ma X, Lu W, Chen B, Zhong D, Huang L, Dong L, Jin C and Zhang Z 2015 Performance change of few layer black phosphorus transistors in ambient *AIP Advances* **5** 107112

[26]    Zhou Q, Chen Q, Tong Y and Wang J 2016 Light-Induced Ambient Degradation of Few-Layer Black Phosphorus: Mechanism and Protection *Angew. Chem. Int. Ed.* **55** 11437–41

[27]    Xu Y, Shi Z, Shi X, Zhang K and Zhang H 2019 Recent progress in black phosphorus and black-phosphorus-analogue materials: properties, synthesis and applications *Nanoscale* **11** 14491–527

[28]    Liu B, Köpf M, Abbas A N, Wang X, Guo Q, Jia Y, Xia F, Weihrich R, Bachhuber F, Pielnhofer F, Wang H, Dhall R, Cronin S B, Ge M, Fang X, Nilges T and Zhou C 2015 Black Arsenic-Phosphorus: Layered Anisotropic Infrared Semiconductors with Highly Tunable Compositions and Properties *Adv. Mater.* **27** 4423–9

[29]    Long M, Gao A, Wang P, Xia H, Ott C, Pan C, Fu Y, Liu E, Chen X, Lu W, Nilges T, Xu J, Wang X, Hu W and Miao F 2017 Room temperature high-detectivity mid-infrared photodetectors based on black arsenic phosphorus *Sci. Adv.* **3** e1700589

[30]    Ozdemir I, Ozaydin H D, Arkin H and Aktürk E 2019 Functionalization of monolayer AsP phases by adatoms: a first-principles study *Mater. Res. Express* **6** 065032

[31]    Li L L, Bacaksiz C, Nakhaee M, Pentcheva R, Peeters F M and Yagmurcukardes M 2020 Single-layer Janus black arsenic-phosphorus (b-AsP): Optical dichroism, anisotropic vibrational, thermal, and elastic properties *Phys. Rev. B* **101** 134102

[32]    Zhang J, Jia S, Kholmanov I, Dong L, Er D, Chen W, Guo H, Jin Z, Shenoy V B, Shi L and Lou J 2017 Janus Monolayer Transition-Metal Dichalcogenides *ACS Nano* **11** 8192–8

[33]    Lu A-Y, Zhu H, Xiao J, Chuu C-P, Han Y, Chiu M-H, Cheng C-C, Yang C-W, Wei K-H, Yang Y, Wang Y, Sokaras D, Nordlund D, Yang P, Muller D A, Chou M-Y, Zhang X and Li L-J 2017 Janus monolayers of transition metal dichalcogenides *Nature Nanotech* **12** 744–9

[34]    Trivedi D B, Turgut G, Qin Y, Sayyad M Y, Hajra D, Howell M, Liu L, Yang S, Patoary N H, Li H, Petrić M M, Meyer M, Kremser M, Barbone M, Soavi G, Stier A V, Müller K, Yang S, Esqueda I S, Zhuang H, Finley J J and Tongay S 2020 Room-Temperature Synthesis of 2D Janus Crystals and their Heterostructures *Adv. Mater.* **32** 2006320

[35]    Yan X, Ke Q and Cai Y 2022 Electronic and optical properties of Janus black arsenic-phosphorus AsP quantum dots under magnetic field *Nanotechnology* **33** 265001

[36]    Akhmerov A R and Beenakker C W J 2008 Boundary conditions for Dirac fermions on a terminated honeycomb lattice *Phys. Rev. B* **77** 085423

[37]    Liu T-H, Chen Y-C, Pao C-W and Chang C-C 2014 Anisotropic thermal conductivity of

MoS $_2$ nanoribbons: Chirality and edge effects *Appl. Phys. Lett.* **104** 201909

[38]  Grujić M M, Ezawa M, Tadić M Ž and Peeters F M 2016 Tunable skewed edges in puckered structures *Phys. Rev. B* **93** 245413

[39]  Ramasubramaniam A and Muniz A R 2014 *Ab initio* studies of thermodynamic and electronic properties of phosphorene nanoribbons *Phys. Rev. B* **90** 085424

[40]  Liu Y, Xu F, Zhang Z, Penev E S and Yakobson B I 2014 Two-Dimensional Mono-Elemental Semiconductor with Electronically Inactive Defects: The Case of Phosphorus *Nano Lett.* **14** 6782–6

[41]  Liang L, Wang J, Lin W, Sumpter B G, Meunier V and Pan M 2014 Electronic Bandgap and Edge Reconstruction in Phosphorene Materials *Nano Lett.* **14** 6400–6

[42]  Ren Y, Liu P, Zhou B, Zhou X and Zhou G 2019 Crystallographic Characterization of Black Phosphorene and its Application in Nanostructures *Phys. Rev. Applied* **12** 064025

[43]  Groth C W, Wimmer M, Akhmerov A R and Waintal X 2014 Kwant: a software package for quantum transport *New J. Phys.* **16** 063065

[44]  Guo H, Lu N, Dai J, Wu X and Zeng X C 2014 Phosphorene Nanoribbons, Phosphorus Nanotubes, and van der Waals Multilayers *J. Phys. Chem. C* **118** 14051–9

[45]  Maity A, Singh A and Sen P 2014 Peierls transition and edge reconstruction in phosphorene nanoribbons

[46]  Carvalho A, Rodin A S and Castro Neto A H 2014 Phosphorene nanoribbons *EPL* **108** 47005